# Sensor Drift Compensation via Olfactory system and Reservoir Computing

ZhengChen Dong
*Kyushu University*
*Information Science and Electrical Engineering*
Fukuoka,Japan
dong@con.ed.kyushu-u.ac.jp

ChenWei Li
*Kyushu University*
*Information Science and Electrical Engineering*
Fukuoka,Japan
li@con.ed.kyushu-u.ac.jp

Takeaki Yajima*
*Kyushu University*
*Information Science and Electrical Engineering*
*Kyushu University*
yajima@ed.kyushu-u.ac.jp

***Abstract*—Despite the promising applications of electronic noses (e-Noses) in medical diagnosis and industrial process control, sensor drift remains a critical challenge that degrades long-term sensing reliability by inducing gradual shifts in sensor responses. Conventional drift compensation methods are typically designed for batch learning and lack the ability to support continuous online learning in non-stationary environments. Although several online drift compensation methods have recently been proposed, they are mainly based on quasi-online mini-batch learning for distribution adaptation, while true sample-wise online learning without buffering remains largely unexplored. To address these issues, this paper proposes a sample-wise online drift compensation method based on spiking neural networks (SNNs) for feature adaptation and spiking reservoir computing (SRC) for classification. By exploiting spike-timing-dependent plasticity (STDP), SNN self-organizes spatiotemporal attractor dynamics for label-free feature adaptation (STDP-FA), and the adapted features are classified by SRC with self-supervised adaptation driven by winner-take-all (WTA) competition. The proposed method addresses various concept drift patterns, including gradual drift, random drift, sensor failures, and abrupt changes. Simulations on a real-world sensor drift dataset demonstrate a clear improvement in classification accuracy over baseline methods.**

## I. INTRODUCTION

With the rapid development of intelligent sensing technologies, e-Noses consisting of chemical gas sensor arrays have been intensively studied for real-world applications including environmental monitoring, industrial process control, medical diagnosis, and safety inspection. In contrast to conventional single-gas sensors that precisely measure the concentration of a specific gas, e-Noses aim to identify odor sources from the response patterns of sensor arrays. Therefore, e-Noses are expected as next-generation edge sensors capable of real-time classification of diverse everyday odors in a bio-inspired manner. However, the long-term reliability of e-Noses remains a major challenge due to sensor drift, a phenomenon in which sensor responses gradually change over time as a result of aging, contamination, and environmental variations. Since e-Noses rely on machine learning to classify sensor response patterns, even slight temporal variations in sensor characteristics can distort the learned feature space and classification boundaries, leading to significant degradation in recognition performance.

Drift compensation for e-Noses can be broadly categorized into two approaches: re-adaptation of the feature space and retraining of the classifier. In both cases, models are typically pre-trained using labeled source data and subsequently updated using target data acquired during operation. Calibration using labeled target data is often adopted in practice because it allows reliable performance recovery [1]-[8]. However, it imposes severe constraints in real-world sensing scenarios. Therefore, semi-supervised approaches that utilize unlabeled data, as well as self-supervised approaches based on internal label assignment, are more desirable in practice. In other words, there is a strong demand for a learning framework that can simultaneously achieve feature adaptation and classifier updating without requiring any ground-truth labels during online operation.

Another major operational constraint in drift compensation is the amount of data required for model adaptation. In many previous studies, batch learning has been adopted by jointly utilizing source data and multiple sets of target data. For example, domain adaptation methods that aim to correct distributional shifts in target data and recover the source data distribution are often performed via batch learning based on unlabeled target data distributions [12]-[17]. However, batch-based approaches require large memory and computational resources to store and process accumulated data. Consequently, their deployment on edge devices, where such resources are severely limited, is highly constrained. Online learning methods that sequentially compensate drift using a small amount of target data have also been proposed [18],[19]. However, they essentially rely on mini-batch buffering and therefore do not fully resolve the above limitations.

In contrast, sample-wise online learning, which performs drift compensation by processing target samples one by one without data accumulation, enables compact implementations and is well suited for edge-device deployment. Nevertheless, existing sample-wise online drift compensation methods remain limited, for example, to classifier-only adaptation [20]-[22], and true sample-wise online adaptation without buffering remains largely unexplored. This limitation arises because most existing approaches are based on learning frameworks that rely on batch statistics or explicit distribution alignment, making stable and immediate adaptation to individual samples fundamentally difficult.

TABLE I. COMPARISON OF METHOD IN REFERENCE

| Ref. | Feature adaptation | Training | | Classification | Training | | Comment |
|---|---|---|---|---|---|---|---|
| | | Labeled data | **Unlabeled data** | | Labeled data | **Unlabeled data** | |
| [1] | Linear (PLS) | batch | – | – | – | – | |
| [2] | Linear (CPCA) | batch | – | k-NN | – | – | |
| [3] | Linear (OSC+PCA) | batch | – | k-NN | – | – | |
| [4] | – | – | – | SVM, KELM, k-NN | batch | – | *Active learning |
| [5] | – | – | – | SVM | batch* | – | *Ensemble |
| [6] | – | – | – | ELM (ODELM) | sample-wise online* | – | *Active learning |
| [7] | Spiking CNN | batch | – | Baysian-SNN | batch | – | |
| [8] | SNN | batch | – | full connect NN | batch | – | |
| [9] | – | – | – | Logistic regression | batch | **batch*** | *Regularization |
| [10] | – | – | – | ELM (DAELM) | batch | **batch** | |
| [11] | CNN (MPC-CNN) | batch | – | FCNN + MMD | batch | **batch** | |
| [12] | Linear (DRCA) | batch | **batch** | RBF-SVM | batch | – | |
| [13] | Geodesic Flow Kernel (WGFK) | batch | **batch** | SVM | batch | – | |
| [14] | ELM (DDR-ELM) | batch | **batch** | ELM-SVM | online | – | |
| [15] | Linear | batch | **batch** | SVM | batch | – | |
| [16] | Linear | batch | **batch** | ELM (SWKELM) | online | **batch** | |
| [17] | Linear | batch | **batch** | FCNN | batch | **batch** | |
| [18] | Linear | – | **mini-batch online*** | RF, PLS | batch | – | *Self-supervised |
| [19] | – | – | – | SVM (CDDA) | mini-batch online* | **mini-batch online*** | *Use recent past data as source |
| [20] | – | – | – | A²INET | batch | **sample-wise online*** | *Self-supervised |
| [21] | – | – | – | ELM (DAELM) | sample-wise online | **sample-wise online*** | *Regularization |
| [22] | Spiking CNN | batch | – | ERSTDP SNN | batch | **sample-wise online*** | *Self-supervised |
| Our work | SNN | – | **sample-wise online*** | Spiking reservoir | batch | **sample-wise online**** | *STDP **Self-supervised |

To address these issues, SNNs with STDP-based local learning are well suited for their unique capability for stable sample-wise online learning. Indeed, Imam and Cleland proposed that an olfactory SNN is capable of rapid online learning and robust recall in noisy environments due to local synaptic plasticity and self-organized attractor dynamics although their work did not explicitly address sensor drift compensation [23]. From a drift compensation perspective, the local and immediate weight updates triggered by individual spike events in STDP would enable systems to continuously adapt to evolving sensory inputs, thereby naturally supporting sample-wise online drift compensation. Building on this concept, we further project the self-organized spatiotemporal patterns in the SNN into a spiking reservoir, which enables fast and stable online classification of the evolving dynamics. The SRC learning is based on the self-supervised adaptation driven by WTA competition. Thus, a combination of STDP-based feature adaptation (STDP-FA) and SRC-based classification provides a feasible alternative solution for sensor drift compensation, particularly in scenarios that require online and incremental learning.

To address the challenges of online sensor drift compensation under continuously evolving sensing conditions, we propose an online self-supervised sensor drift compensation framework using SNN combined with SRC. To the best of our knowledge, the use of SNNs jointly with reservoir computing for sensor drift compensation has not yet been reported in the existing literature, and this framework provides a practical solution for long-term online drift adaptation without requiring labeled target data. The main novelties of this work can be summarized as follows:

- We propose a continuously adaptive sample-wise online drift compensation scheme in which a SNN and a reservoir-based classifier incrementally learn from unlabeled inputs. By exploiting event-driven spiking dynamics and local plasticity, the SNN self-organizes drift-resilient spatiotemporal dynamics, while the reservoir classifier adapts its readout through WTA-driven self-supervised learning. This enables stable drift compensation without buffering or offline retraining as sensor characteristics progressively evolve.
- Extensive simulations validate the robustness and efficiency of our framework under heterogeneous drift

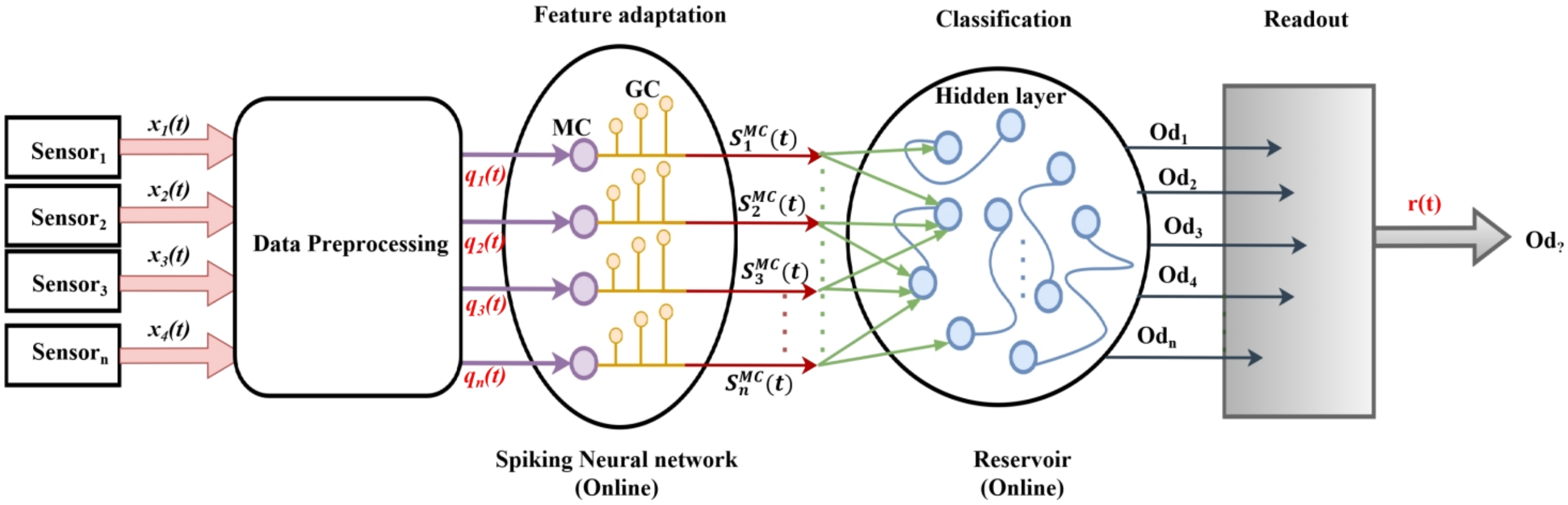

Fig. 1: Strategy of spiking neural network with reservoir.

mechanisms, including gradual drift, random drift, sensor failures, and abrupt changes. These results suggest that the proposed method generalizes well beyond a single drift assumption and can handle multiple realistic drift modes within a unified online compensation pipeline.

- Under comparable experimental settings on public datasets commonly adopted in prior work, our method delivers higher recognition accuracy than representative DA methods and ensemble classifiers, highlighting clear advantages over conventional drift-compensation pipelines.

## II. Related Work

### A. Background of Drift Compensation

Early studies on sensor drift primarily focused on offline calibration and statistical correction techniques (blue color in Table 1). Haugen *et al.* first addressed temporal drift through periodic calibration strategies for solid-state gas sensors, establishing a foundational framework for drift handling in practical deployments [24]. Subsequently, signal preprocessing methods were introduced to suppress drift-related variations, including Orthogonal Signal Correction (OSC), which removes drift components orthogonal to the target information [3]. In parallel, subspace-based approaches such as Common Principal Component Analysis (CPCA) were proposed to model shared variance structures across time and compensate for drift in a global statistical manner [2]. Online active learning paradigms incorporating mixed-kernel classifiers have also been proposed to selectively update models with informative samples [4]. Moving beyond pure signal correction, Vergara *et al.* demonstrated that classifier ensemble techniques could improve robustness against drift by leveraging model diversity, representing an early attempt to address drift at the decision level rather than the sensor level [5].

Following early calibration and signal-correction approaches, subsequent studies increasingly addressed sensor drift at the classification level, aiming to enhance decision robustness directly (green color in Table 1). Zhang *et al.* further integrated extreme learning machines, proposing DAELM to transfer discriminative knowledge across drift domains efficiently [10]. Despite these advances, classifier-based methods remain sensitive to label uncertainty, sample selection bias, and accumulated model errors, which can lead to degraded performance under severe or long-term drift.

As sensor drift was increasingly recognized as a distribution shift problem, domain adaptation (DA) methods were introduced to align data distributions across different drift stages (orange color in Table 1). Early studies applied semi-supervised domain adaptation to electronic noses, leveraging unlabeled target-domain samples to reduce distribution mismatch while preserving classification performance [13]. More recent studies focused on subspace transfer and adversarial learning frameworks, enabling more robust feature alignment between source and target domains under complex drift patterns [15].

Despite their effectiveness, most DA-based classifier methods still rely on batch-wise domain alignment, which restricts their deployment to offline or weakly online settings and limits their capability for continuous real-time adaptation. Several studies have explored online drift compensation by updating the recognition model during operation. Ma *et al.* proposed an online framework that continuously adapts the ELM classifier to track time-varying sensor responses under drift [21]. Martinelli *et al.* developed an adaptive classifier inspired by artificial immune systems, leveraging immune-like memory and adaptation to mitigate drift without explicit domain alignment [20]. More recently, Mei *et al.* demonstrated that local plasticity in spiking networks can support continual update of classification in gradually changing environments, providing a neuromorphic route to online drift robustness [22].

### B. Motivation of the Proposed Research

Motivated by recent advances in the neuromorphic olfactory system, SNNs provide a principled foundation for addressing sensor drift under continuous operation. Imam and Cleland demonstrated rapid online learning and robust recall in a neuromorphic olfactory circuit through local synaptic plasticity, highlighting that spike-based models can adapt continuously to evolving input statistics without batch re-training or optimization [23]. In the context of gas sensing, Huo et al. showed that SNNs can support few-shot class-incremental gas recognition, providing evidence that spike-based architectures can update decision-relevant representations as new conditions or classes emerge over time [25]. Separately, Xue et al. developed a biomimetic SNN-based electronic nose for mixed-gas recognition, illustrating the effectiveness of spike-driven temporal encoding for complex chemical patterns [26]. Therefore, an SNN can be exploited as an online adaptive representation model, in which information is conveyed by discrete spikes that arrive over time, and learning can proceed incrementally as these spike events arrive. Therefore, the SNN enables continuous online feature extraction. Moreover, since the reservoir module is also implemented as a spiking, event-driven network, we integrate the SNN and the reservoir to realize an online domain-adaptive transformation.

## III. System Modeling

### A. Mechanism and Formulation

As shown in Fig. 1, we propose an online, self-supervised sensor-drift compensation framework that couples an olfactory SNN with reservoir computing. The sensor data are preprocessed before being fed into the SNN. The system runs continuously on sequential sensor outputs, where the preprocessed data are converted into spike events, processed by the SNN, and immediately forwarded to the reservoir. Both modules are updated on the fly as new data arrive, enabling drift compensation without requiring labeled target samples.

The olfactory SNN consists of a recurrent two-layer network: a layer of Mitral cells (MCs) and a layer of granule cells (GCs). Raw sensor vectors are encoded into sparse spatiotemporal spike patterns through neuron dynamics and local online synaptic plasticity. Specifically, excitatory plasticity is implemented on the MC-to-GC synapses, which are strengthened to reinforce stable, odor-relevant activation pathways. In parallel, inhibitory plasticity is implemented on the GC-to-MC feedback projections, where the learned GC-to-MC inhibition adaptively regulates MC firing by suppressing unstable co-activations and controlling runaway excitation. This coordinated MC-to-GC excitatory adaptation and GC-to-MC inhibitory adaptation maintains an appropriate excitation-inhibition balance. As a result, the SNN provides drift-resilient spatiotemporal representations suitable for the subsequent reservoir computing.

On top of the SNN output spike pattern, the reservoir computing module performs classification with online learning. The reservoir projects the incoming spikes into a high-dimensional dynamical state space via fixed recurrent dynamics, while the trainable readout is rapidly updated to track representation shifts induced by drift. Concretely, the readout weights are adapted online through a spike-driven STDP rule coupled with winner-take-all (WTA) competition. The WTA mechanism provides internal label assignments for incoming patterns, which in turn guide self-supervised adaptation through STDP. The reservoir computing classifier improves separability among odor classes, and the continuously adapted readout maintains high classification performance as the input spike patterns change over time.

### B. Spike-Based Olfactory SNNs

#### a) Event-Driven Discrete Spike Representation

In spike-based olfactory SNNs, information is conveyed by asynchronous, event-driven spike trains rather than continuous-valued signals. For neuron *i*, the emitted spikes are represented as a point process:

$$S_i(t) = \sum_k \delta(t - t_i^k) \tag{1}$$

Where $t_i^k$ denotes the time of *k*-th spike emitted by neuron *i*, and $\delta(.)$ is the Dirac delta. This formulation makes explicit that neural communication is discrete in time and sparse in activity, with state updates and synaptic interactions being triggered by the arrival of events.

For numerical simulation and streaming implementations, time is typically discretized as $t = n\Delta t$, and spikes are encoded as binary events:

$$S_i[n] \in \{0,1\}, \;\; S_i[n] = \mathbb{1}(V_i[n] \geq V_{th}) \tag{2}$$

where $\mathbb{1}(.)$ is the indicator function, $V_i[n]$ is the membrane potential, and the $V_{th}$ is the firing threshold. Under this event-based representation, downstream synaptic and learning dynamics can be implemented in an event-triggered manner: state variables are incremented at spike times and otherwise degrade according to their intrinsic time constants. This property is particularly suitable for olfactory sensing scenarios.

#### b) Sensor-to-Spike Encoding

Let the raw response of a *D*-channel sensor array at time *t* be denoted by:

$$x(t) = [x_1(t), \dots x_D(t)]^T \in R^D \tag{3}$$

We first enforce sparsity to select sensors whose outputs are not weak or non-informative, using top-k gating to yield a sparsified vector $G(t) \in R^D$:

$$G(t) = T_k(x(t)), \;\; [T_k(v)]_d = \begin{cases} v_d & d \in TopK(v,k) \\ 0 & otherwise \end{cases} \tag{4}$$

For each channel, we map the gated response into [0,1] using channel-wise bounds $(x_d^{min}, x_d^{max})$:

$$u_d(t) = \frac{G_d(t) - x_d^{min}}{x_d^{max} - x_d^{min}} \tag{5}$$

*d* denotes the *d*-th channel, and we then quantize the bounded value into a compact integer code with *B* levels:

$$q_d(t) = \left\lfloor (B-1)u_d(t) + \frac{1}{2} \right\rfloor \tag{6}$$

Collecting all channels gives $q_d(t) \in \{0,\dots,B-1\}^D$, which is both bounded and sparse due to the prior gating.

#### c) Apical Dendrite event generation

The MC is modeled with an apical dendrite (AD) that generates a cycle-locked initiation event and a soma compartment that emits the output spike. The soma spike is generated only if an AD event has occurred in the current gamma cycle and if the soma is not suppressed by GC-induced inhibitory gating (blocking). A gamma cycle denotes one oscillatory period of the olfactory network rhythm, which provides a natural time window for organizing spiking activity. Within each gamma cycle of duration $T_\gamma$, we define a permissive phase window $W_\gamma$ during which the MC apical dendrite is allowed to initiate at most one spike event. This phase gating is implemented by a binary indicator $g_\gamma(t) \in \{0,1\}$, where $g_\gamma(t)$=1 if $t \in W_\gamma$ and $g_\gamma(t)$=0 otherwise. The AD converts the sensory drive into a single spike-initiation event whose latency encodes input strength. Note that $W_\gamma$ denotes a strict sub-interval within the nth gamma cycle, i.e, the permissive phase occupies only a fraction of the cycle rather than the entire period.

Let $x_m$ denote the sniff-wise coded sensor drive delivered to the *m*th MC, and let $t_m^{AD}$ be the first time in the permissive window at which the AD crosses threshold. We represent the AD event as:

$$e_m^{AD}(t) = \mathbb{1}\{t = t_m^{AD}\} \quad (7)$$

$$t_m^{AD} = min\{t: x_p(t) = 1, a_m(t) \geq \theta_{AD}\} \quad (8)$$

where $a_m(t)$ is an AD activation state driven by $x_m$ during the permissive epoch. $x_p(t)$ is a binary phase-gating indicator of the sniff gamma cycle that equals 1 when time *t* falls within the permitted phase for event generation, and 0 otherwise. $\theta_{AD}$ is the AD event threshold.

*d) Online training of SNNs*

To enable online, self-supervised adaptation in the olfactory SNN, synaptic efficacies are updated using local spike-driven plasticity rules. We employ complementary plasticity mechanisms: (i) Excitatory synaptic plasticity from MCs to GCs is implemented via spike-timing-dependent plasticity (STDP), (ii) Inhibitory plasticity from lateral GCs to MCs stabilizes network activity and maintains sparsity. Importantly, $t_m^{AD}$ provides a cycle-local reference time used by plasticity: it is the target alignment point for learning inhibitory timing (via GC-to-MC blocking/release), while MC output spikes generated around this event serve as the presynaptic events for excitatory plasticity on MC to GC synapses. The MC soma emits an output spike only when the AD has initiated and the soma is not within a GC-induced blocking window. Let $B_m(t) \in \{0,1\}$ denote the aggregate inhibitory blocking state on the *m*th MC. The soma spike time $t_m^{MC}$ is defined by

$$t_m^{MC} = min\{t: x_p(t) = 1 \;\&\; t \geq t_m^{AD} \;\&\; B_m(t) = 0\} \quad (9)$$

$$s_m^{MC}(t) = 1\,\{t = t_m^{MC}\} \quad (10)$$

$x_p(t)$ denote the permissive phase indicator of the sniff–gamma cycle. This makes explicit the mechanism required for inhibitory plasticity: GC feedback modulates MC spike latency by blocking propagation until a learned release time[23]. Each GC integrates a sparse set of delayed MC spikes and emits (at most) one spike per gamma cycle,

$$v_g(t) = \sum_{k \in P(g)} w_{gk}^E s_k^{MC}(t - d_{gk}) \quad (11)$$

$$s_g^{GC}(t) = 1\{v_g(t) \geq \theta_{GC}\} \quad (12)$$

$P(g)$ denotes presynaptic MCs for the *g*th GC. $w_{gk}^E$ denotes the excitatory synaptic weight from the *k*th MC to the *g*th GC. $d_{gk}$ is a fixed synaptic delay of the spike train of the *k*th MC arriving at the *g*th GC.. These variables are sufficient for excitatory plasticity: when a GC spikes at $t = t_g^{GC}$, the relative timing between $t_g^{GC}$ and its contributing presynaptic MC spikes determines potentiation versus depression on $w_{gk}^E$.

A GC spike triggers an inhibitory blocking window on the local MC soma whose duration $\Delta B_{gm}$ is the inhibitory strength learned online. For a GC spike at time $t_g^{GC}$, define the blocking indicator $b_{gm}(t)$:

$$b_{gm}(t) = 1\{t_g^{GC} \leq t \leq t_g^{GC} + \Delta B_{gm}\} \quad (13)$$

The corresponding release time $t_{gm}^R$ is

$$t_{gm}^R = t_g^{GC} + \Delta B_{gm} \quad (14)$$

Inhibitory plasticity updates $\Delta B_{gm}$ so that $t_{gm}^R$ aligns with the AD initiation timing $t_m^{AD}$, thereby learning an odor-specific inhibitory timing profile that controls MC spike latencies across gamma cycles.

## C. Reservoir Computing

*a) State Update of Spiking Reservoir*

At each time step, the reservoir receives the current input spikes and recurrent feedback from the previous reservoir spikes:

$$u(t) = W_{in}z(t) + W_{rec}r(t-1) \quad (15)$$

$u(t)$ denotes the reservoir's interal state at time *t*. It is composed of two terms:(i) an input-driven component $W_{in}z(t)$; and (ii) a recurrent component $W_{rec}r(t-1)$, which feeds back the previous reservoir state $r(t-1)$ via the recurrent connectivity $W_{rec}$.

Using a standard discrete-time LIF abstraction consistent with a leak rate $\beta$:

$$v_{res}(t) = \beta v_{res}(t-1) + u(t),$$
$$r(t) = \Theta(v_t - \vartheta_r) \quad (16)$$

$\vartheta_r$ is the reservoir threshold, and $\Theta$ is the Heaviside step function. $r(t)$ is the spike function of reservoir.

*b) Online Classification of Odor Sequence*

We adopt a two-phase learning protocol composed of offline training and online training. Let $\zeta(s)$ denote the set of samples observed at drift stage $s$, and the offline corpus can be defined:

$$\zeta_{off} = \cup_{s \in s_{off}} \zeta(s) \quad (17)$$

In the online stage, each incoming stages $s \in S_{on}$ is processed in a adapt manner under the fixed decoding rule, which is defined by the WTA operator, $\psi$. Let $y(x; W_{out}) \in \mathbb{R}^c$ be the readout score vector for sample $x$, where $W_{out}$ is the readout weight matrix and $C$ is the number of classes. The WTA operator is defined as following:

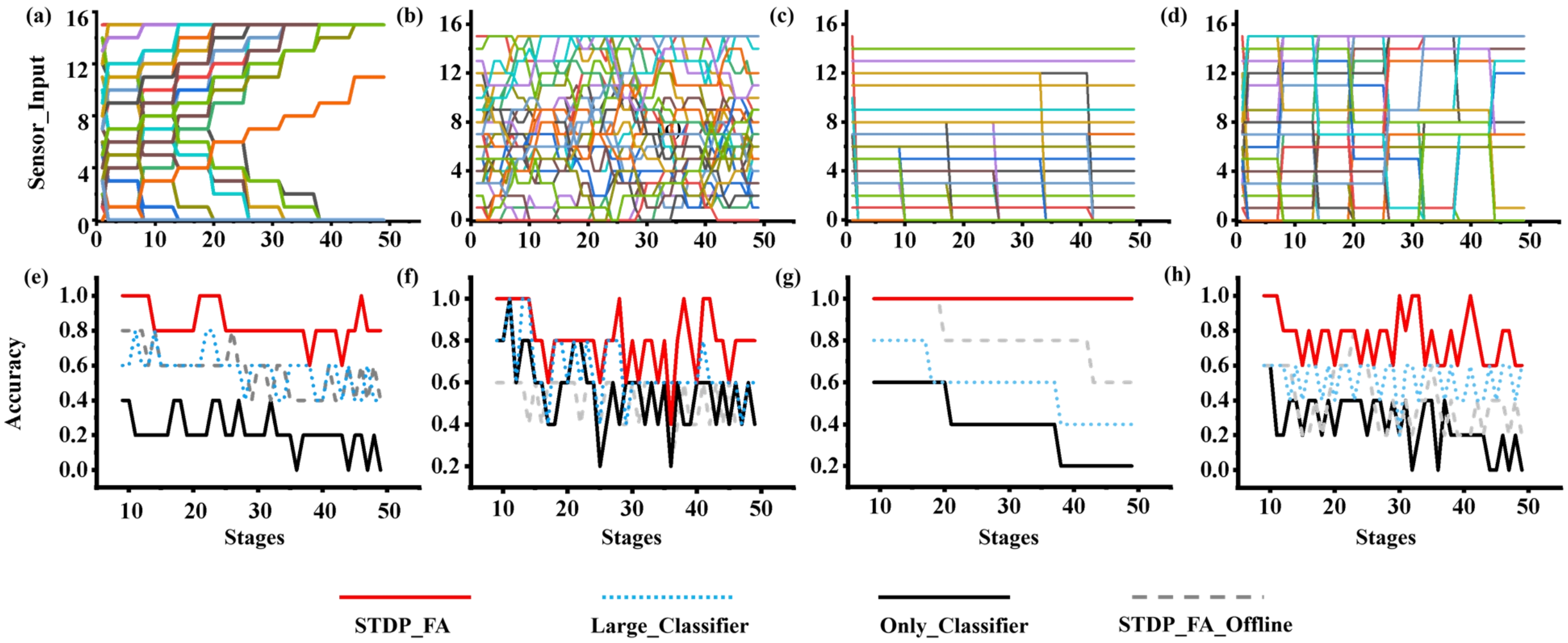

Figure. 2: Drift scenarios and performance comparison under ablation settings. (a)-(d):Drift trajectories of a representative odor sample under four drift regimes: (a) gradual drift, (b) random drift, (c)sensor failure, and (d) abrupt change;(e)-(h):Gas identification accuracy under the corresponding drift regimes in (a)-(d).

$$\psi(y) = arg_{c\in[\{1\ldots C\}}\, max\, y_c \tag{18}$$

Then perform self-supervised online adaptation on $\zeta(s)$,yielding the updated readout weights:

$$W_{out}^{post}=W_{out}^{pre} + \sum \Delta\, \mathrm{W}_{out}(\mathrm{x};\eta_{on}) \tag{19}$$

Where $\eta_{on}$ is the online learning rate and $\Delta \mathrm{W}_{out}(\mathrm{x};\eta_{on})$ denotes the cumulative the WTA-triggered STDP increments produced while processing sample *x*. Finally we compute the post-adaptation accuracy using the same equation:

$$ACC_s^{post} = ACC(W_{out}^{post};\zeta(s);\psi) \tag{20}$$

## IV. Results Analysis

In this section, we evaluate the proposed online, self-supervised drift compensation framework under four representative drift scenarios and quantify its compensation effectiveness across drift stages. Specifically, we construct controlled drift simulations spanning distinct non-stationary modes, thereby covering common degradation patterns observed in practical gas-sensing deployments. For each drift type, sequential stages are generated to emulate long-term sensor evolution, and the model is operated in a strictly online manner: samples are processed in temporal order, with performance assessed after adaptation to characterize both immediate robustness and incremental compensation capability. We use the e-Nose dataset reported in [5]. Following the original data organization, we treat the early stages as source data for initialization and then train online the remaining stages as unlabeled target data to emulate real deployment conditions. The evaluation reports stage-wise recognition accuracy, providing a comprehensive view of how effectively the method tracks drift while preserving odor-discriminative structure.

### A. Multi-Type Drift Compensation

As shown in Fig. 2, we show the gas-identification accuracy observed under different drift conditions for four configurations: STDP_FA, Large_Classifier, Only_Classifier and STDP_FA_Offline. In "STDP_FA", the FA denotes the feature extractor. STDP indicates that both the olfactory system and the reservoir are included, and STDP is applied in both parts. "STDP_FA_Offline" is the offline version of STDP_FA, where the SNN and reservoir are trained with STDP only in the online stage.

"Only_Classifier" is a baseline that uses only the reservoir without the SNN. "Large_Classifier" also means the reservoir without the SNN, where the number of neurons that would have been used in the SNN is instead added to the reservoir to test the effect of increasing node capacity when using only the reservoir. Each stage consists of one sample for each of the five odors . Stages 1-50 are ordered chronologically from the earliest to the most recent measurements, thereby capturing long-term temporal drift caused by prolonged sensor usage. In Fig. 2, (a) – (d) show the drift trajectory of a single odor only, provided as an illustrative example.

In addition, we include a large-scale reservoir curve to illustrate the case where the SNN module is removed and a larger number of neurons is allocated to the reservoir. This curve serves as a proxy for conventional baselines that rely solely on a reservoir-based classifier [27]. In the simulation, the input is a 36-channel gas-sensor array, and the corresponding MC representation therefore consists of 36 MC units (one per sensor channel).

For the EPL-style instantiation, the number of GCs is determined by allocating 5 GCs per MC per odor. Since we simulate five gas classes, the total GC population is 5× 36 × 5. The simulation spans a total of 50 stages, where the progression of stages emulates cumulative temporal evolution and the sensor outputs become increasingly affected by drift. In our protocol, offline learning is performed on Stages 1 – 10 for both the SNN and the reservoir module, while Stages 11 – 50 operate in an online manner. Offline learning is performed on Stages 1-10 for both the SNN and the reservoir modules by jointly training on all samples from these stages in a single batch. In contrast, Stages 11-50 operate online: once a new stage arrives, the SNN and the reservoir immediately perform incremental updates using the newly observed sensor responses from that stage, enabling continuous self-supervised adaptation to progressive drift. Specifically, when a new drifted sensor sample arrives, the SNN first processes the incoming sensor vector, performs online synaptic adaptation, and encodes it into a sparse spatiotemporal spike sequence. This spike sequence is then forwarded to the reservoir, which conducts online STDP-WTA learning on its readout. Immediately after this online update, the system outputs a gas-identification decision for the same input, thereby enabling continual learning and inference under stage-wise drift evolution.

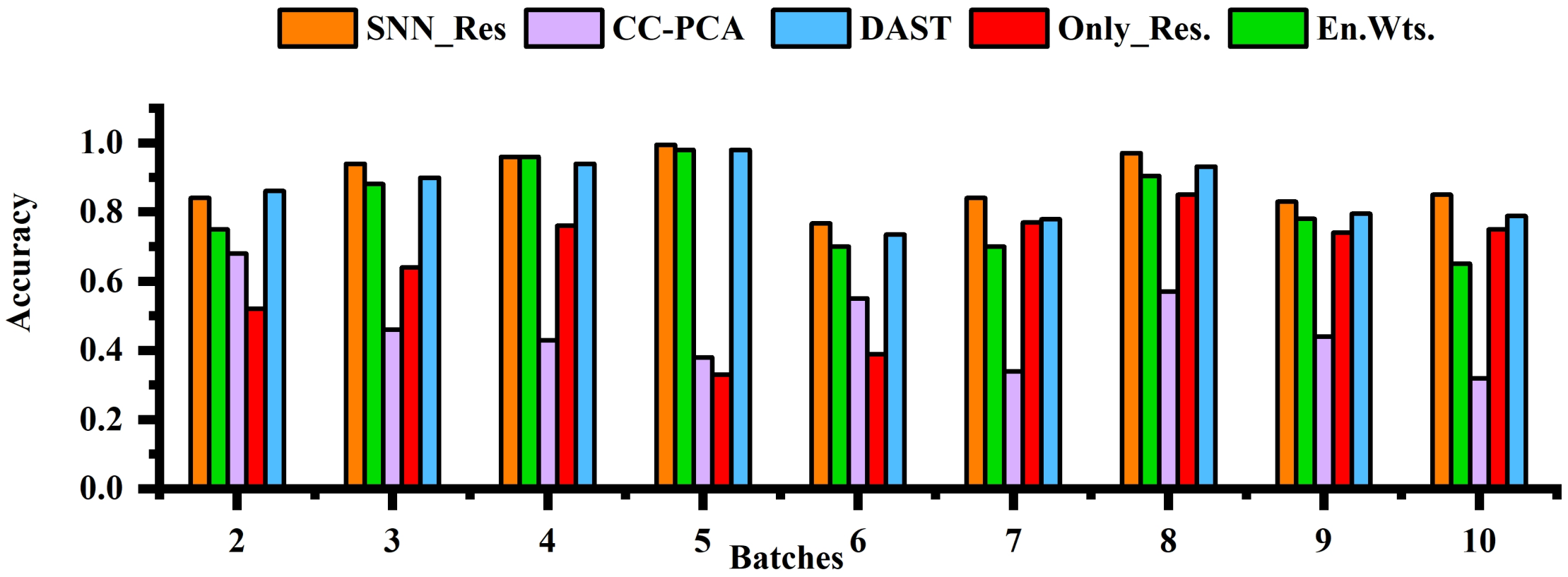

Fig. 3: Classification Accuracy of Different Method

As can be observed, the sensor failure case corresponds to a failure mode in which one or more sensor channels progressively deteriorate and, after a certain time point, can no longer provide valid or usable responses.

In particular, the large-scale reservoir uses a number of nodes equal to the sum of (i) the reservoir size in our main setting (300 nodes) and (ii) the total number of neurons that would otherwise be allocated to the SNN module (900 GCs and 36 MCs). The purpose of this curve is to provide a capacity-matched comparison: under the same total neuron budget, it contrasts the accuracy of a reservoir-only classifier against the proposed SNN-reservoir framework. As evidenced in Fig. 2, the proposed online SNN-reservoir framework achieves the most effective drift compensation, consistently maintaining the highest gas-identification accuracy across drift conditions compared with configurations that remove the SNN module and/or disable online learning.

### B. Dataset Comparison

We additionally evaluated our method on the dataset used by Vergara et al. [5], which contains sensor responses to six different gases measured at multiple concentration levels and spans 36 months of recordings. The data are organized into ten batches (Batch 1 – Batch 10), corresponding to different time periods. Batch 1 corresponds to sensor measurements collected at Month 1 and Month 2. Batch 2 includes Months 3, 4, 8, 9, and 10. Batch 3 includes Months 11, 12, and 13. Batch 4 includes Months 14 and 15. Batch 5 contains Month 16 only. Batch 6 includes Months 17, 18, 19, and 20. Batch 7 contains Month 21 only. Batch 8 includes Months 22 and 23. Batch 9 includes Months 24 and 30. Batch 10 contains Month 36 only. Here, Month 1 denotes the earliest sampling time, whereas Month 36 corresponds to the most recent sampling time. Since the number of sensor channels is 128, we adjusted the number of MCs accordingly and then conducted comparative experiments against conventional baselines.. As summarized in Fig. 3, we compare our approach with the ensemble classifier reported in [5], the Component Correction by Principal Component Analysis (CC-PCA) method, and a standard offline domain adaptation method DAST[15]. In the Only_Res (Only reservoir), we include a Large_Classifier control setting to explicitly demonstrate the limitation of removing the spike-based olfactory SNN network and only use reservoir computing to classify the odors. Under a comparable total neuron budget, this setting discards the SNN front-end and relies solely on an enlarged reservoir, thereby highlighting the performance degradation that arises when drift compensation is attempted without the SNN online spike-based representation learning. Similarly, following the evaluation protocol and baselines in [5], we report the best-performing gas for each method to enable a fair comparison. As shown in Fig. 3, the proposed approach consistently outperforms the ensemble classifier in [5], CC-PCA, DAST and the Only_Res. This indicates that the proposed method achieves strong performance not only across multiple types of drift data, but also on existing benchmark datasets.

## V. Conclusion

We propose an online, self-supervised sensor-drift compensation architecture that couples a SNN-based feature adaptation with a reservoir computing classification. Both the SNN and the reservoir readout learn online, enabling continuous operation on streaming sensor inputs and continual weight updates as drift evolves. In contrast to conventional domain adaptation pipelines, where the transformation is typically estimated in an offline manner and is difficult to update reliably in real time, our framework provides a fully online alternative that adapts directly to non-stationary sensing conditions without requiring labeled target data. Simulation results demonstrate that the proposed method not only achieves consistently stable drift compensation across multiple drift types, but also delivers strong recognition performance on established benchmark datasets. In particular, when evaluated on the dataset used in prior work, our approach yields notably improved accuracy especially in the middle and later stages, providing a clear classification accuracy improvement compared with representative conventional baselines. Future work will focus on deploying the proposed SNN-reservoir framework on neuromorphic hardware to validate its real-time performance, computational efficiency, and energy characteristics in practical sensing systems, and to conduct comprehensive hardware-in-the-loop evaluations under long-term drift.

Authorship and Version Note. This manuscript is a revised preprint based on the paper accepted for presentation at ICANN 2026. The present version includes Chenwei Li and Takeaki Yajima as co-authors in recognition of their substantial contributions to the work. Their omission from the conference submission resulted from an administrative error during the submission process. All authors have reviewed and approved the authorship and content of this preprint. This clarification applies only to the present preprint and does not alter the ICANN 2026/Springer version of record.